\documentclass[conference]{IEEEtran}
\IEEEoverridecommandlockouts

\usepackage{cite}
\usepackage{algorithm}
\usepackage{algorithmic}
\usepackage{textcomp}
\usepackage{amsmath}
\usepackage{amssymb}
\usepackage{mathrsfs}
\usepackage{epsfig}
\usepackage{epsf}
\usepackage{theorem}
\usepackage[active]{srcltx}
\usepackage{enumerate}
\usepackage{latexsym,calc,dsfont,array,pgf,setspace}
\usepackage{acro}
\usepackage{tikz}
\usetikzlibrary{calc}
\usepackage{pifont}
\usepackage{subfigure}
\usepackage{epstopdf}
\usepackage{amsfonts}
\usepackage{float}
\usepackage{xcolor}
\usepackage{afterpage}
\usepackage[inkscapeformat=png]{svg}
\usepackage{graphicx}
\usepackage[acronym,shortcuts]{glossaries}
\usepackage{xcolor}
\usepackage{titlesec}
\usepackage{url}
\usepackage{orcidlink}

\newacronym{FH}{FH}{frequency-hopping}
\newacronym{JRC}{JRC}{joint radar and communication}
\newacronym{MIMO}{MIMO}{multiple-input and multiple-output}
\newacronym{SIMO}{SIMO}{single-input and multiple-output}
\newacronym{DFRC}{DFRC}{dual-function radar communications}
\newacronym{AF}{AF}{ambiguity function}
\newacronym{PRI}{PRI}{pulse repetition interval}
\newacronym{PRF}{PRF}{pulse repetition frequency}
\newacronym{PPH}{PPH}{polynomial-phase hopping}
\newacronym{QSM}{QSM}{quadrature spatial modulation}
\newacronym{RIS}{RIS}{reconfigurable intelligent surface}
\newacronym{ISAC}{ISAC}{integrated sensing and communications}
\newacronym{AWGN}{AWGN}{additive white Gaussian noise}
\newacronym{CPM}{CPM}{continuous phase modulation}
\newacronym{ECCM}{ECCM}{electronic counter-countermeasures}
\newacronym{RFA}{RFA}{random frequency agility}
\newacronym{RPA}{RPA}{random \ac{PRI} agility}
\newacronym{RFPA}{RFPA}{random frequency and \ac{PRI} agility}
\newacronym{QAM}{QAM}{quadrature amplitude modulation}
\newacronym{SINR}{SINR}{signal-to-interference plus noise ratio}
\newacronym{SNR}{SNR}{signal-to-noise ratio}
\newacronym{ASK}{ASK}{amplitude shift keying}
\newacronym{PSK}{PSK}{phase shift keying}
\newacronym{ML}{ML}{maximum likelihood}
\newacronym{RIP}{RIP}{restricted isometry property}
\newacronym{OMP}{OMP}{orthogonal matching pursuit}
\newacronym{CIR}{CIR}{channel impulse response}
\newacronym{TDD}{TDD}{time division duplex}
\newacronym{CRKG}{CRKG}{channel reciprocity-based key generation}
\newacronym{FCM}{FCM}{Fuzzy C-means}
\newacronym{PIRS}{PIRS}{polar code-based information scheme}
\newacronym{BER}{BER}{bit error rate}
\newacronym{CRC}{CRC}{cyclic redundancy check}
\newacronym{BDR}{BDR}{bit disagreement rate}
\newacronym{SQ}{SQ}{scalar quantization}
\newacronym{VQ}{VQ}{vector quantization}
\newacronym{6G}{6G}{sixth-generation}
\newacronym{AI}{AI}{artificial intelligence}
\newacronym{V2V}{V2V}{vehicle to vehicle}
\newacronym{IM}{IM}{index modulation}
\newacronym{SM}{SM}{spatial modulation}
\newacronym{SIM}{SIM}{spatial index modulation}
\newacronym{PLS}{PLS}{physical layer security}
\newacronym{AoA}{AoA}{angle-of-arrival}
\newacronym{JCAS}{JCAS}{joint communication and sensing}
\newacronym{SotA}{SotA}{state-of-the-art}
\newacronym{FHCS}{FHCS}{frequency hopping code selection}
\newacronym{FHCSK}{FHCSK}{frequency hopping code-shift keying}
\newacronym{PH}{PH}{phase}
\newacronym{TX}{TX}{transmit}
\newacronym{AMP}{AMP}{amplitude}

\begin{document}

\title{A Secure ISAC Waveform Design Framework via Random Frequency and PRI Agility}

\author{
\IEEEauthorblockN{Ali Khandan Boroujeni,
        Hyeon Seok Rou,
        Ghazal Bagheri,
        Giuseppe Thadeu Freitas de Abreu,\\
        Stefan Köpsell,
        Kuranage Roche Rayan Ranasinghe,
        and Rafael F. Schaefer}
\thanks{A. K. Boroujeni, S. Köpsell, and R. F. Schaefer are with Barkhausen Institut and Technische Universit\"at Dresden, 01067 Dresden, Germany (emails: ali.khandanboroujeni@barkhauseninstitut.org; \{stefan.koepsell, rafael.schaefer\}@tu-dresden.de). H. S. Rou, G. T. F. de Abreu, and K. R. R. Ranasinghe are with the School of Computer Science and Engineering, Constructor University, Bremen, Germany (emails: \{gabreu, kranasinghe\}@constructor.university). G. Bagheri is with Technische Universit\"at Dresden, 01187 Dresden, Germany (email: ghazal.bagheri@tu-dresden.de). \textbf{An extended version of this work has been submitted to IEEE Transactions on Information Forensics and Security.}}
}

\maketitle

\begin{abstract} 
This paper presents a novel framework for enhancing the security, data rate, and sensing performance of \ac{ISAC} systems. 
We employ a random frequency and \ac{PRI} agility (\acs{RFPA}) method for the waveform design, where the necessary random sequences are governed by shared secrets. %
These secrets, which can be pre-shared or generated via channel reciprocity, obfuscate critical radar parameters like Doppler frequency and pulse start times, thereby significantly impeding the ability to perform reconnaissance from a passive adversary without the secret key. 
To further introduce enhanced data throughput, we also introduce a hybrid information embedding scheme that integrates \ac{ASK}, \ac{PSK}, \ac{IM}, and \ac{SM}, for which a low-complexity sparse-matched filter receiver is proposed for accurate decoding with practical complexity. 
Finally, the excellent range-velocity resolution and clutter suppression of the proposed waveform are analyzed via the \ac{AF}. 
\end{abstract}


\glsresetall

\section{Introduction}
The evolution toward \ac{6G} wireless systems is driving the integration of communication and sensing into a unified \ac{ISAC} architecture \cite{Liu_JSC22, Wei_ITJ23}. 
By sharing spectrum and hardware resources, \ac{ISAC} can achieve significant efficiency gains and support emerging applications such as autonomous transportation, industrial automation, augmented reality, and environmental monitoring \cite{ISACMarket2023, WangITJ2022}. 
In these scenarios, a vehicle may simultaneously exchange data with the network and neighboring nodes while employing the same signals for mapping its environment and detecting potential hazards.

This dual functionality, while highly attractive, also introduces new security challenges. 
The signals designed for sensing may be intercepted by passive adversaries who, without transmitting, can exploit reflected echoes to track user locations, infer activities, and reconstruct sensitive environmental information \cite{KaiqianISAC2023}. 
Safeguarding both the communication data and the sensing information is therefore essential for the reliable and secure deployment of \ac{ISAC} \cite{su2023security}, further especially in anticipation of the pervasive connectivity and \ac{AI}-integration of \ac{6G} networks \cite{ZhangNetwork2024,ChenWCom2023}.

While substantial progress has been made for secure communication links through cryptography and \ac{PLS} \cite{ylianttila6GSec2020,MucchiOJCS2021, Bagheri_JCN25} by using approaches such as jamming and waveform design \cite{SuTWC_2021,Rou_WCL24}, ensuring the confidentiality of the sensing function remains more difficult and typically require additional energy and/or hardware such as \acp{RIS} or artificial noise \cite{Rexhepi_Asilomar2025,zou2024securing}. 
Therefore, to efficiently realize secure \ac{ISAC}, the waveform \textit{itself} must guarantee reliable target detection, high-rate communication, and resilience against unauthorized analysis.

In light of the above, \ac{FH} signaling is an increasingly popular approach, as its inherent low probability of intercept and robustness to jamming, long emphasized in military contexts, align naturally with these requirements \cite{WuTWC2022}.
Nevertheless, adapting \ac{FH} signaling to \ac{ISAC} has highlighted a fundamental trade-off among security, throughput, and sensing performance. 
Approaches that embed information in the hop selection or permutation using \ac{IM} and \ac{SM} \cite{BaxterTSP_2022, Boroujeni_JCS24, HassanienRC_2017} can enhance data rates but often degrade the radar \ac{AF}, generating high sidelobes and necessitating receivers of prohibitive complexity. 
In contrast, methods that randomize the \ac{PRI} to improve radar resolution \cite{LongTAES_2021} typically focus on sensing aspects while neglecting integration with high-throughput communication. 
As a result, a comprehensive framework that jointly addresses secure signaling, reliable sensing, and efficient communication is still lacking.

To address this gap, this article proposes a secure and efficient \ac{FH}-\ac{ISAC} waveform that leverages \acf{RFPA} by introducing secret key-controlled randomness in both the carrier frequency and \ac{PRI}. This approach renders the signal structure inherently unpredictable to passive adversaries. Specifically, the \ac{PRI} obscures the delay domain, significantly complicating attempts by a passive malicious radar to estimate target range, while the \ac{RFA} obscures the Doppler domain, impeding accurate target velocity estimation. Simultaneously, this dual-layer randomization enhances the physical layer security of the communication messages, effectively mitigating the risk of eavesdropping.
Furthermore, to meet communication demands, a hybrid embedding scheme is developed which integrates \ac{ASK}, \ac{PSK}, \ac{IM}, and \ac{SM}, accompanied by a novel low-complexity matched filter receiver exploiting the frequency-domain sparsity of \ac{FH} signals.

\section{System and Signal Model}
\label{Preliminaries}
\subsection{ISAC Wiretap Channel Model}

\begin{figure}[t]
\centering
\includegraphics[width=0.8\columnwidth]{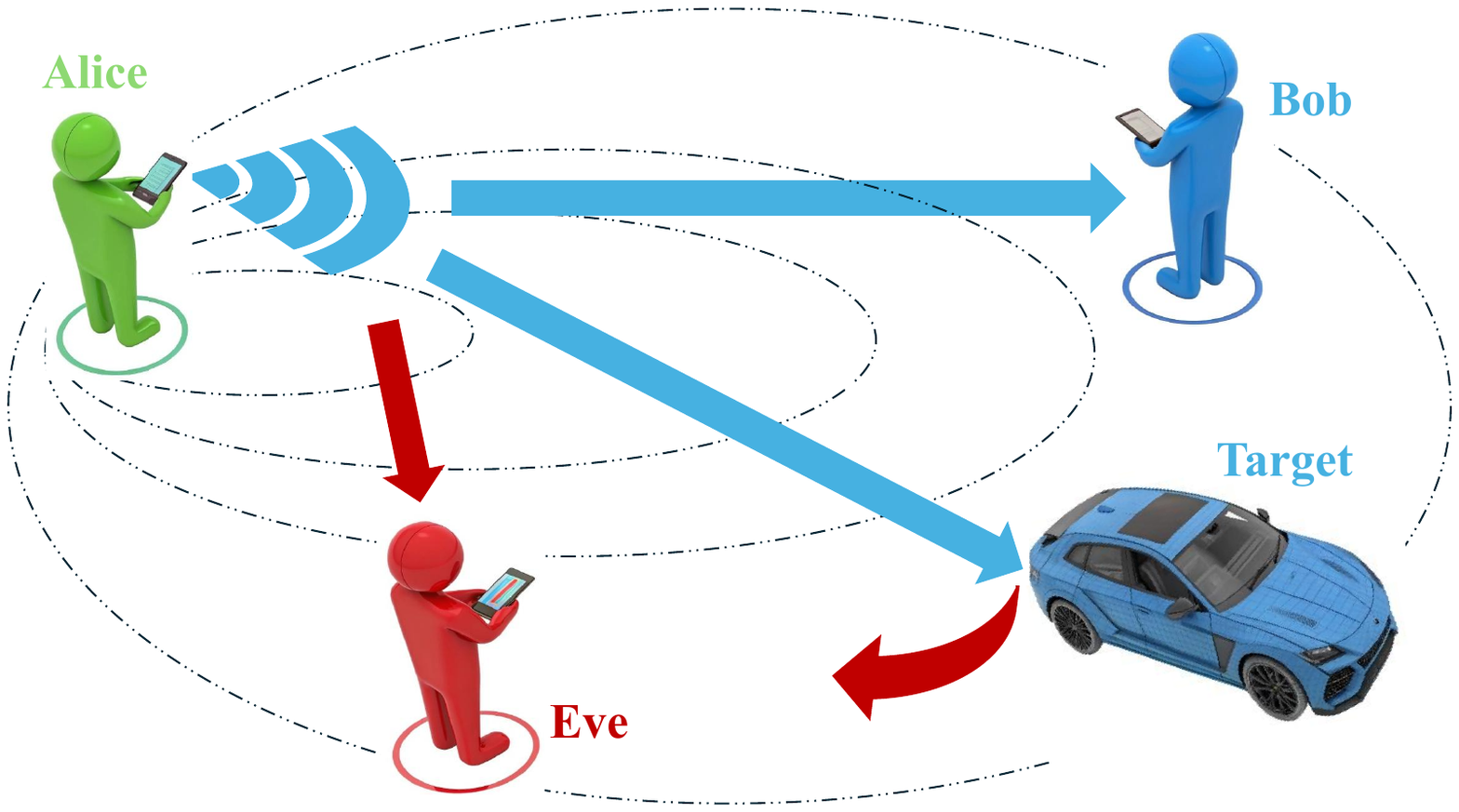}
\caption{System model for secure \ac{RFPA}-\ac{FH}-\ac{ISAC}, where a passive Eve eavesdrops on the communication and sensing links.}
\label{fig:system_model}
\end{figure}

As illustrated in Fig.~\ref{fig:system_model}, consider two legitimate pre-authenticated communication partners: Alice and Bob, equipped with linear arrays of $M$ transmit and $N$ receive antennas, separated by distances $d_{T}$ and $d_{R}$, respectively. 
Eve, a passive eavesdropper equipped with communication and sensing receivers, seeks to intercept Alice's transmissions. 
 
Alice embeds information into her \ac{ISAC} \ac{FH} waveform, transmitting it towards Bob and a target (possibly Bob himself), in order to jointly estimate the range and velocity of the targets. 
Both Bob and Eve attempt to extract the embedded information from the received signal, modeled as \cite{LiuSPM_2023}
\begin{align}
\label{eq:receive_signal}
\mathbf {r}(t;l) &= \mathbf {H}_l\mathbf {x}(t;l) + \mathbf {v}(t;l) \in \mathbb {C}^{N},\\[1ex]
\mathbf {r}^{(e)}(t;l) &= \mathbf {H}^{(e)}_l\mathbf {x}(t;l) + \mathbf {v}^{(e)}(t;l) \in \mathbb {C}^{N},
\end{align}
where $\mathbf {H}_l\in \mathbb {C}^{N\times M}$ and $\mathbf {H}^{(e)}_l \in \mathbb {C}^{N\times M}$ denote the flat-fading channel matrices between Alice and Bob and Alice and Eve, respectively, with elements $h_{i,j}$ and $h_{i,j}^{(e)}$ following $\mathcal{CN}(0,1)$, which remain constant during the $l^{\text{th}}$ \ac{FH} pulse;
the transmit signal is denoted by $\mathbf {x}(t;l) \in \mathbb {C}^{M}$ at time $t$ during the $l^{\text{th}}$ pulse, while $\mathbf {v}(t;l), \mathbf {v}^{(e)}(t;l)\in \mathbb {C}^{N}$ denote \ac{AWGN} with elements $v_{i,j}\sim \mathcal{CN}(0,{\sigma^2_{\mathbf{v}}})$ and $v_{i,j}^{(e)}\sim \mathcal{CN}(0,\sigma^2_{\mathbf{v}^{(e)}})$, where ${\sigma^2_{\mathbf{v}}}$ and $\sigma^2_{\mathbf{v}^{(e)}}$ represent the respective noise powers at Bob and Eve. 

The quasi-static Rayleigh fading channel matrix $\mathbf{H}$ is assumed perfectly known at the receiver but unknown to the transmitter. 
Because Eve remains passive, her location and channel are unknown to the legitimate parties, which prevents the use of countermeasures such as beamforming, artificial noise injection, or constructive interference. 

Given the above scenario, the objective of this work is to design an \ac{ISAC} \ac{FH} waveform that enhances resilience against Eve through \ac{PLS}, while simultaneously improving radar estimation accuracy and communication performance.
\vspace{-1.5em}

\subsection{Frequency Hopping MIMO Signal Model}

In a conventional \ac{FH} system, each pulse of duration $\tau$ seconds is divided into $Q$ chips of length $\Delta_t \triangleq \tau/Q$, and the waveform transmitted by the $m^{\text{th}}$ antenna is given by \cite{BaxterTSP_2022}
\begin{equation}
\label{eq:pulse_signal_model}
x_m(t) = \sum ^{Q-1}_{q=0} e^{{\imath
}2\pi c_{m,q}\Delta _f t }  \Pi_q(t) = \sum ^{Q-1}_{q=0}h_{m,q}(t)\Pi _q(t),
\end{equation}
where $h_{m,q}(t)\triangleq \exp\{{\imath}
2\pi c_{m,q} \Delta_f t\}$ denotes the \ac{FH} signal at the $q^{\text{th}}$ chip and the $m^{\text{th}}$ antenna, $c_{m,q}\in\mathcal{K}\triangleq \{0, 1, \ldots, K-1\}$ is the hop code, and $\Pi_q(t)$ is the window function $\Pi(t-q\Delta_t)$ defined as \vspace{2ex}
\begin{equation}
\label{eq:window_function}
\Pi(t)\triangleq \Big\lbrace 
\begin{array}{ll}1, & 0 \leq t \leq \Delta _t,\\ 0, & {\text{otherwise}}. 
\end{array} 
\end{equation}

The main \ac{FH} design parameters in an \ac{ISAC} system include $\Delta_f$, $\Delta_t$, $K$, $M$, and $Q$, which must ensure spectral confinement within the allocated bandwidth. 
The number of transmit antennas $M$ is bounded by $\tfrac{K}{Q} \leq M \leq KQ$, with the case $K = MQ$ yielding orthogonal chip cross-correlations and low sidelobes, and the upper limit $M \leq KQ$ corresponds to the maximum number of orthogonal waveforms but results in higher sidelobes due to frequent hop reuse. 
Furthermore, orthogonality is achieved when $\Delta_t = 1/\Delta_f$ and $K\Delta_f \leq BW$, where $BW$ is the available radar bandwidth, so to preserve orthogonality, chips within each pulse must satisfy
\begin{equation}
\label{eq:orthogonality}
c_{m,q} \ne {c_{m^{\prime}, q}}, \quad \forall q, m \ne {m^{\prime}}.
\end{equation}
Finally, while not required for basic radar operation, the condition $M \leq K$ becomes necessary for information embedding schemes to enable reliable detection at communication receivers with matched filters, as will be shown later.

\section{Secure \ac{RFPA} Waveform Design}
\label{Waveform-Design}
Building upon the conventional \ac{FH} structure, we introduce a new waveform that incorporates random agility in both frequency and time to enhance security and sensing performance.

\subsection{Proposed Secure \ac{RFPA}-\ac{FH}-\ac{ISAC} Waveform}
\label{RFPA-FH-ISAC}
To counter passive eavesdropping and improve radar ambiguity characteristics, we introduce randomness into the pulse structure itself by randomizing the starting time and the base carrier frequency of each pulse. 
This approach, dubbed \acf{RFPA}, makes the signal structure unpredictable to an adversary without knowledge of the underlying random sequences.

The proposed generalized waveform for the $m^{\text{th}}$ transmit antenna is formulated as \vspace{1ex}
\begin{equation}
\label{eq:RFPA_waveform}
\begin{aligned}
x_m(t) = & \sum^{L-1}_{l=0}  \sum^{Q-1}_{q=0} a_{m,q}^{(l)} e^{{\imath}\Omega_{m,q}^{(l)}} e^ {{\imath}2\pi (f_l+c_{m,q}^{(l)}\Delta_f)(t-lT_p-T_l)} \\ 
&\times \Pi \left ({t-q\Delta_t-lT_p-T_l}  \right ),
\end{aligned}
\end{equation}
where $a_{m,q}^{(l)}$ and $\Omega_{m,q}^{(l)}$ are amplitude and phase modulation symbols. 

\begin{figure}[t]
\centering
\includegraphics[width=\columnwidth]{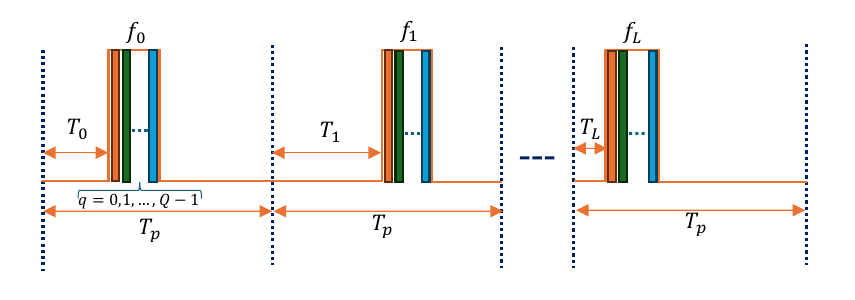}
\vspace{-3ex}
\caption{Illustration of the proposed secure \ac{RFPA}-\ac{FH}-\ac{ISAC} waveform.}
\label{fig:signal_model}
\vspace{-2ex}
\end{figure}

Then, the modulated output signal at Alice can be expressed as an $M \times 1$ vector across $M$ transmit antennas at time $t$ in pulse $l$, given by
\begin{eqnarray}
\label{eq:vector_RFPA_waveform}
\mathbf{x}(t;l) = \sum ^{Q-1}_{q=0}\text{diag}\big({\mathbf a}_{q}^{(l)}\odot e^{{\imath}\boldsymbol{\Omega }_{q}^{(l)}}\big)&& \\[-1ex]
&& \hspace{-26ex}\times \exp \left\lbrace {\imath}2\pi  (f_l \mathbf{1}_M + {\mathbf c}_{q}^{(l)} \Delta _f) (t-T_l)\right\rbrace \Pi _q(t-T_l),\nonumber 
\end{eqnarray}
where ${\mathbf a}_{q}^{(l)}$ and $\boldsymbol{\Omega}_{q}^{(l)}$ denote the amplitude and constant \ac{PSK} phase rotation vectors for the $M$ waveforms, respectively. 

To strengthen physical-layer security against Eve, the parameters $T_l$ and $f_l$ are quantized as
$T_l =  \Delta_{T_L} \times \phi_{T_l}$ and $f_l =  \Delta_{f_L} \times \phi_{f_l}$,
where $\Delta_{T_L} \triangleq Q \Delta_t = \tau$ and $\Delta_{f_L} \triangleq K\Delta_f$ are constant quantization values. The integers $\phi_{T_l}$ and $\phi_{f_l}$ are randomly drawn from the sets $\varphi_{T_l}\triangleq\{ 0, 1, \ldots, \Phi_{T_l} -1 \}$ and $\varphi_{f_l}\triangleq\{ 0, 1, \ldots, \Phi_{f_l} -1 \}$, with $\Phi_{T_l} \triangleq (T_p /\Delta_{T_L})-1$ and $\Phi_{f_l} \triangleq BW/(K\Delta_f)$, both designed as powers of two. 
Therefore, key security features are the agility parameters: $T_l$, a random time offset for the $l$-th pulse within its \ac{PRI}, and $f_l$, a random frequency offset for the $l$-th pulse. 
These parameters are governed by secret integer sequences, $\mathbf{\Gamma}_T$ and $\mathbf{\Gamma}_f$, shared between Alice and Bob. 
In practice, these secrets can be pre-shared or generated on-the-fly using \ac{CRKG} methods, ensuring that only the legitimate parties can correctly process the signal timing and frequency structure.

\vspace{-1ex}
\subsection{Ambiguity Function Analysis}
\label{subsec:AF_proposed_waveform}
The \ac{AF} is a critical tool for evaluating a waveform's ability to resolve targets in range and Doppler.

Consider a target at $(\tau,\nu,f)$, where $\tau$ represents the delay associated with the target's range, $\nu$ denotes the Doppler frequency of the target, and $f$ indicates the normalized spatial frequency, defined as $f\triangleq 2\pi \frac{{d_{R}}}{\lambda} \sin \theta$, where $\theta$ denotes the angle of the target and $\lambda$ represents the wavelength.
When attempting to capture this target signal using a matched filter with assumed parameters $(\tau',\nu',f')$, the \ac{MIMO} radar \ac{AF} can be characterized as
\begin{equation}
\label{eq:ambiguity_function}
\chi (\tau , \nu , f, f^{\prime}) \triangleq \sum\limits _{m=0}^{M-1}\sum\limits _{m^{\prime}=0}^{M-1} \chi_{m,m^{\prime}}(\tau , \nu) e^{{\imath}2\pi (fm-f^{\prime}m^{\prime})\gamma}, 
\end{equation}
where $\chi_{m,m^{\prime}}(\tau , \nu)$ is the cross-ambiguity function given by
\begin{equation}
\label{eq:cross_ambiguity_function}
\chi_{m,m^{\prime}}(\tau, \nu) \triangleq \int_{-\infty}^{+\infty}x_m(t)x_{m'}^{*}(t+\tau)e^{{\imath}2\pi \nu t}{\rm d}t.
\end{equation}

By substituting our proposed \ac{RFPA} waveform from eq. \eqref{eq:RFPA_waveform} into the definition of the cross-ambiguity function, and some rigorous reformulation omitted in this article, the full \ac{AF} is derived analytically as
\begin{eqnarray}
\label{eq:RFPA_AF(6)}
\chi(\tau,\nu,f,f^{\prime})=\!\!\sum_{m=0}^{M-1}\!\sum_{m^{\prime}=0}^{M-1}\!\sum_{l=0}^{L-1}\! \sum_{l'=0}^{L-1}\!\sum_{q=0}^{Q-1}\!  \sum_{q'=0}^{Q-1}e^{{\imath}2\pi (mf-m^{\prime}f^{\prime})\gamma}&&\\
&&\hspace{-48ex}\quad\times\frac{e^{{\imath}(\Omega_{m,q}^{(l)} - \Omega_{m',q'}^{(l')})} a_{m,q}^{(l)} a_{m',q'}^{(l')}e^{\beta_2}\left( e^{\alpha_2 \beta_1} - e^{\alpha_2 \alpha_1} \right)}{\alpha_2},\nonumber\hspace{-3ex}
\end{eqnarray}

\noindent where the auxiliary terms $\alpha_1, \alpha_2, \beta_1, \beta_2$ defined intrinsically with $t_l \triangleq l T_p + T_l$ and $t_{l'} \triangleq l' T_p + T_{l'}$ are given by 
\begin{align}
\alpha_1 &= \max\big((q'-q)\Delta_t+t_{l'}-t_l-\tau, 0\big), \\
\alpha_2 &= {\imath}2\pi \big( (f_l+c_{m,q}^{(l)}\Delta_f) - (f_{l'}+c_{m',q'}^{(l')}\Delta_f) + \nu \big), \\[0.5ex]
\beta_1  &= \min\big((q'-q +1)\Delta_t+t_{l'}-t_l-\tau, \Delta_t\big), \\
\beta_2  &= {\imath}2\pi \big( (f_l+c_{m,q}^{(l)}\Delta_f)q\Delta_t  \\
         &\hspace{0.5em} + \nu(q\Delta_t+t_l) (f_{l'}+c_{m',q'}^{(l')}\Delta_f)(q\Delta_t+t_l-t_{l'}+\tau)  \big). \nonumber
\end{align}

\section{Information Embedding and Receiver Design}
\label{sec:Information-Embedding-Schemes}

In this section, we illustrate how the proposed framework enables information embedding within radar signals.
First, a hybrid scheme that combines multiple embedding strategies to boost data rates is presented, accompanied by a receiver design for Bob, assuming perfect synchronization and knowledge of the frequency hops, chip interval, \ac{FH} step, and bandwidth.

\subsection{Proposed Transmit Signal Design}
\label{sub:Sparse_Receiver_Hybrid}

In \ac{ISAC} scenarios, radar pulses often allocate significant time for target echoes during the \ac{PRI}, leaving limited opportunity for concurrent data transmission. 
This restriction leads to low communication rates, which can be alleviated using index and spatial modulation to exploit available frequency hops and antenna allocations. 
Further combining these with conventional modulation schemes (e.g., \ac{QAM}) enables higher rates without altering radar operating parameters.

First, the transmitted signal of Alice is redefined as
\begin{align}
\mathbf{x}_{\text{hyb}} (t;l)\! = \sum ^{Q-\!1}_{q=0}\text{diag}\left({\mathbf a}_{q}^{(l)}\!\odot \!e^{{\imath}\boldsymbol{\Omega }_{q}^{(l)}}\!\right) e^{{\imath}2\pi {{\mathbf P}_{q}^{(l)}{\mathbf S}_{q}^{(l)}} {\mathbf d}\Delta _f t} \Pi_q(t\!-\!T_l), \nonumber \\[-3ex]
\label{eq:hyb_transmit_waveform}
\end{align}

\vspace{-1.5ex}
\noindent where ${\mathbf a}_{q}^{(l)}$ and $\boldsymbol{\Omega}_{q}^{(l)}$ are amplitude and phase symbol vectors with entries drawn from $\mathcal{C}_{ASK}$ and $\mathcal{C}_{PSK}$, and the matrix ${\mathbf S}_{q}^{(l)}$ selects chip indices (index modulation), while ${\mathbf P}_{q}^{(l)}$ permutes them across antennas (spatial modulation) -- together establishing the relation ${\mathbf c}_{q}^{(l)} = {\mathbf P}_{q}^{(l)}{\mathbf S}_{q}^{(l)}\mathbf d$ in eq. \eqref{eq:vector_RFPA_waveform}.

\textbf{Phase Embedding:}  
When power efficiency is prioritized, information can be embedded solely in chip phases using PSK. The transmit signal reduces to
\begin{equation}
\label{eq:PSK_embedding}
\mathbf{x}_{\text{ph}}(t;l) \!=\! \sum ^{Q-1}_{q=0}\!\text{diag}\!\left(e^{{\imath}\boldsymbol{\Omega }_{q}^{(l)}}\right) e^{{\imath}2\pi (f_l \mathbf{1}_M + {\mathbf c}_{q}^{(l)}\Delta _f) (t-T_l)} \Pi _q(t-T_l),
\end{equation}
\\
where ${\mathbf a}_{q}^{(l)}=\mathbf{1}_M$. The matrices ${\mathbf P}_{q}^{(l)}$ and ${\mathbf S}_{q}^{(l)}$ are predefined and shared, thus conveying no additional information.

\textbf{Amplitude Embedding:}  
For scenarios emphasizing simplicity and cost efficiency, amplitude modulation (\ac{ASK}) may be applied, yielding the transmit signal
\begin{align}
\mathbf{x}_{\text{amp}}(t;l) = \sum ^{Q-1}_{q=0}\text{diag}\{\boldsymbol{a}_{q}^{(l)}\} e^{{\imath}2\pi (f_l \mathbf{1}_M + {\mathbf c}_{q}^{(l)}\Delta _f) (t-T_l)}
 \Pi _q(t-T_l), \nonumber \\[-2ex] \label{eq:AM_embedding}
\end{align}

\vspace{-2ex}

\noindent where ${\mathbf a}_{q}^{(l)}$ contains amplitude symbols from $\mathcal{C}_{ASK}$.

\textbf{Spatial Index Modulation:}  
Alternatively, spatial modulation embeds information in antenna and index mappings
\begin{equation}
\label{eq:sim_embedding}
\mathbf{x}_{\text{sim}}(t;l) = \sum ^{Q-1}_{q=0} e^{{\imath}2\pi (f_l \mathbf{1}_M + {\mathbf c}_{q}^{(l)}\Delta _f) (t-T_l)} \Pi _q(t-T_l),
\end{equation}
where the matrices ${\mathbf S}_{q}^{(l)}$ and ${\mathbf P}_{q}^{(l)}$ act as selection and permutation matrices in index modulation and spatial modulation, respectively, contributing to the relation ${\mathbf c}_{q}^{(l)}={\mathbf P}_{q}^{(l)}{\mathbf S}_{q}^{(l)}\mathbf d$.

\subsection{Proposed Receiver Design}

Given the transmit signal of Alice, as described in the previous section, which may consist of any of the hybrid embedding methods in eqs. \eqref{eq:PSK_embedding}-\eqref{eq:sim_embedding}, Bob's received signal under an \ac{AWGN} channel is given by
\begin{equation}
\label{eq:hyb_received_signal}
\mathbf{r}_{\text{typ}}(t;l) = \mathbf{H}_l\mathbf{x}_{\text{typ}}(t;l) + \mathbf w(t;l).
\end{equation}

Assuming perfect knowledge of the channel, the transmit signal vector can be approximated simply by 
\begin{eqnarray}
 \hat{\mathbf{x}}_{\text{typ}}(t;l) \hspace{-4ex}&&= {\mathbf{H}_l^{\dagger}} \mathbf{r}_{\text{typ}}(t;l) 
 { \;\approx}\; \mathbf{x}_{\text{hyb}}(t;l) + {\mathbf{H}_l^{\dagger}} \mathbf w(t;l)\nonumber\\
 && =\mathbf{\Psi}_l\hat{\mathbf{s}}_{\text{type}}(t;l)+ {\mathbf{H}_l^{\dagger}} \mathbf w(t;l),
 \label{eq:estimated_transmit_signal_hyb}
\end{eqnarray}
where $\hat{\mathbf{s}}_{\text{typ}}(t;l)$ are sparse signals $\hat{\mathbf{x}}_{\text{typ}}(t;l)$ projected onto the Fourier transform basis $\mathbf{\Psi}_l$, and the subscript ``typ'' denotes the specific type of signal being transmitted.

Given the above, Bob must then estimate $\hat{\mathbf a}_{q}^{(l)}$, $\hat{\boldsymbol{\Omega}}_{q}^{(l)}$, $\hat{\mathbf S}_{q}^{(l)}$, and $\hat{\mathbf P}_{q}^{(l)}$ from $\hat{\mathbf{x}}_{\text{hyb}}(t;l)$, for which different strategies are required depending on the embedding scheme, as elaborated in the following.

\textbf{Phase Embedding:} For the estimation of \ac{PSK} embedded information, Bob applies matched filtering to maximize \ac{SNR}.
First, the vector of $K$ available \ac{FH} waveforms is obtained as
\begin{eqnarray}
\label{eq:available_FH_waveforms_PSK}
\mathbf {h}(t;l) = e^{{\imath}2\pi (f_l \mathbf{1}_K + \textbf {d}\Delta _f (t-T_l))}&& \\ 
&&\hspace{-26ex} = e^{{\imath}2\pi\left[f_l (t-T_l),\;  (f_l +\Delta _f) (t-T_l),\; \dots \;,\;  (f_l + (K-1)\Delta _f) (t-T_l)\right]^T}. \nonumber
\end{eqnarray} 

Then, Bob computes $\tilde{\mathbf h}_q(t;l)={\mathbf P}_{q}^{(l)}{\mathbf S}_{q}^{(l)}\mathbf h(t;l)$ and performs the matched filter as to the \ac{FH} chips as
\begin{equation}
\label{eq:matched_filtering_PSK}
 \mathbf{\gamma}^{(l)}_{q} = \int _{T_l + q\Delta _t}^{T_l + (q+1)\Delta _t}\left(\mathbf{1}_M\cdot\hat{\mathbf{x}}_{\text{ph}}(t;l)\right)\tilde{\mathbf h}_{q}^{*}(t;l){\rm d}t,  
\end{equation} 
from which the phase estimates are obtained as
\begin{equation}
\label{eq:PSK_symbols_estimation_PSK}
\hat{\boldsymbol{\Omega}}_{q}^{(l)} = \angle {\boldsymbol{{\gamma}}_{q}^{(l)}}.
\end{equation} 
\textbf{Amplitude Embedding:} Similarly, amplitude information can be extracted by simply performing a matched filter as
\begin{equation}
\label{eq:matched_filtering_AM}
\hat{\mathbf a}_{q}^{(l)} = \int _{T_l + q\Delta _t}^{T_l + (q+1)\Delta _t}
\left(\mathbf{1}_M \cdot \hat{\mathbf x}_{\text{amp}}(t;l)\right)\tilde{\mathbf h}_{q}^{*}(t;l){\rm d}t.  
\end{equation} 

\textbf{Spatial Index Modulation:} For \ac{SIM}, decoding requires estimating ${\mathbf S}_{q}^{(l)}$ and ${\mathbf P}_{q}^{(l)}$, i.e., the index and spatial mappings, from the received signal.
The optimal \ac{ML} receiver must solve
\begin{equation}
\label{eq:ML}
\left\lbrace \hat{\mathbf{c}}_q^{(l)}\right\rbrace _{k=0}^{K-1} = \mathop {\arg \min } \limits _{\lbrace{\mathbf{c}}_q^{(l)}\rbrace } \big\Vert \mathbf{x}_{\text{sim},q}(t;l) - \hat{\mathbf{x}}_{\text{sim},q}(t;l)\big\Vert _2^2,
\end{equation}
but a naive search-based approach becomes computationally infeasible due to the combinatorial nature of the discrete search space of the index modulation codewords \cite{BaxterTSP_2022,Rou_TWC24}. 

Therefore, to reduce complexity, we exploit the sparsity of the signal in the frequency domain, reformulating it as
\begin{equation}
\label{eq:sparse_ML}
\begin{aligned}
\left\lbrace \hat{\mathbf{c}}_q^{(l)}\right\rbrace _{k=0}^{K-1} = & \mathop {\arg \min } \limits _{\lbrace{\mathbf{c}}_q^{(l)}\rbrace } \big\Vert \mathbf{x}_{\text{sim},q}(t;l) - \mathbf{\Psi}_l\hat{\mathbf{s}}_{\text{sim},q}(t;l)\big\Vert _2^2\\
& \quad\quad \mathrm{s.t.}\; \big\Vert \hat{\mathbf{s}}_{\text{sim},q}(t;l) \big\Vert _{l_1}=M,
\end{aligned}
\end{equation}
where $\mathbf{\Psi}_l$ is the Fourier dictionary. 

Since the received signal is exactly $1$-sparse in frequency per chip and antenna, we employ the \ac{OMP} \cite{DavenportTIT_2010,CaiTIT_2011}.

When a hybrid modulation scheme is employed, the receiver on Bob's side must extract information from \ac{ASK}, \ac{PSK}, index, and spatial modulations by jointly estimating \(\hat{\mathbf a}_{q}^{(l)}\), \(\hat{\boldsymbol{\Omega}}_{q}^{(l)}\), \(\hat{\mathbf S}_{q}^{(l)}\), and \(\hat{\mathbf P}_{q}^{(l)}\), respectively, from the received signal \(\hat{\mathbf{x}}_{\text{hyb}}(t;l)\).
To achieve this, a novel receiver combining \ac{OMP} and matched filtering is proposed as Algorithm \ref{alg:Sparse_Matched_Filter_Receiver}, enabling efficient extraction of information from the noisy hybrid-modulated signal.

\begin{algorithm}[h]
\caption{Sparse MF Receiver Design for Hybrid Mod.}
\label{alg:Sparse_Matched_Filter_Receiver}
\scriptsize
\begin{algorithmic}[1]
\renewcommand{\algorithmicrequire}{\textbf{Input:}}
\renewcommand{\algorithmicensure}{\textbf{Output:}}
\REQUIRE $\mathbf{r}_{\text{hyb}}(t;l)$ and $\mathbf{H}_l$
\ENSURE  $\hat{\mathbf a}_{q}^{(l)}$, $\hat{\boldsymbol{\Omega}}_{q}^{(l)}$, $\hat{\mathbf S}_{q}^{(l)}$ and $\hat{\mathbf P}_{q}^{(l)}$
\FOR {each pulse $l=0$ to $L-1$}
\STATE Calculate $\mathbf {h}(t;l)$ based on \eqref{eq:available_FH_waveforms_PSK}.
\FOR{each sub-pulse $q=0$ to $Q-1$}
    \STATE Calculate $\hat{\mathbf{x}}_{\text{hyb}}(t;l) = {\mathbf{H}_l^{\dagger}} \mathbf{r}_{\text{hyb}}(t;l)$.
    \FOR{each antenna element $m=0$ to $M-1$}
        \STATE $\rho = (\hat{s}_{m,q}(:;l))$
        \STATE $\ell_{\rho} = \text{length}(\rho)$.
        \STATE Let
        $\mathbf{\Psi}(i,j) \triangleq \exp\{-{\imath} 2 \pi i j / \ell_{\rho}\},  \forall i, j = 0, \dots, \ell_{\rho}-1.$
        \STATE Select atom: $\hat{c}_{m,q}^{(l)} = (\arg\max_i |\langle \mathbf{\Psi}_i, \rho \rangle|)\times \frac{f_s}{\ell_{\rho}}$.
    \ENDFOR
    \STATE Form $\hat{\mathbf{c}}_{q}^{(l)}=\big[\hat{c}_{0,q}^{(l)}, \hat{c}_{1,q}^{(l)}, \cdots, \hat{c}_{M-1,q}^{(l)}\big]$
    \STATE Compute $\hat{{\mathbf S}}_{q}^{(l)}$ and $\hat{{\mathbf P}}_{q}^{(l)}$ so that $\hat{{\mathbf c}}_{q}^{(l)}=\hat{{\mathbf P}}_{q}^{(l)}\hat{{\mathbf S}}_{q}^{(l)}\mathbf d$.
    \STATE Define $\tilde{\mathbf h}_q(t;l)\triangleq \hat{{\mathbf P}}_{q}^{(l)} \hat{{\mathbf S}}_{q}^{(l)} \mathbf h(t;l)$
    \STATE Using Matched filtering \\
    $\boldsymbol{\gamma}^{(l)}_{q} = \int _{T_l + q\Delta _t}^{T_l + (q+1)\Delta _t} \left(\mathbf{1}_M\cdot\hat{\mathbf{x}}_{\text{hyb}}(t;l)\right)\tilde{\mathbf h}_{q}^{*}(t;l){\rm d}t$,
    \STATE $\hat{\mathbf a}_{q}^{(l)} = \mid{\boldsymbol{\gamma} ^{(l)}_{q}}\mid$ and $\hat{\boldsymbol{\Omega}}_{q}^{(l)} = \angle{\boldsymbol{\gamma} ^{(l)}_{q}}$
\ENDFOR
\ENDFOR
\end{algorithmic} 
\end{algorithm}

\section{Performance Analysis}
\label{Results}
Finally, in this section, we evaluate the performance of our proposed framework through numerical simulations and analysis.
The system parameters are set as follows unless otherwise stated: carrier frequency \(f_c = 10\) GHz, bandwidth \(BW = 200\) MHz, \(K = 10\) available hops, pulse duration \(\tau = 2\) µs, \(Q=10\) chips, \(M=N=8\) antennas, \ac{ASK} size \(J_{ASK} = 2\), and \ac{PSK} size { \(J_{PSK} = 4\)}.
\subsection{Achievable Bit Rate}
The achievable bit rate denotes the maximum data transmission rate over the channel, expressed in bits per unit time, and is determined by PRF, $M$, $K$, $Q$, $J_{PSK}$, and $J_{ASK}$, which for the total hybrid scheme is obtained as
\begin{equation}
\label{eq:Hybrid_Achievable_Bit_Rate}
R_{\text{hyb}} = R_{\text{sim}} + \text{PRF} \times Q \times M  \left\lfloor\log_2 (J_{ASK} J_{PSK})\right\rfloor.  
\end{equation}

\begin{figure}[h]
\centering
\includegraphics[width=\columnwidth]{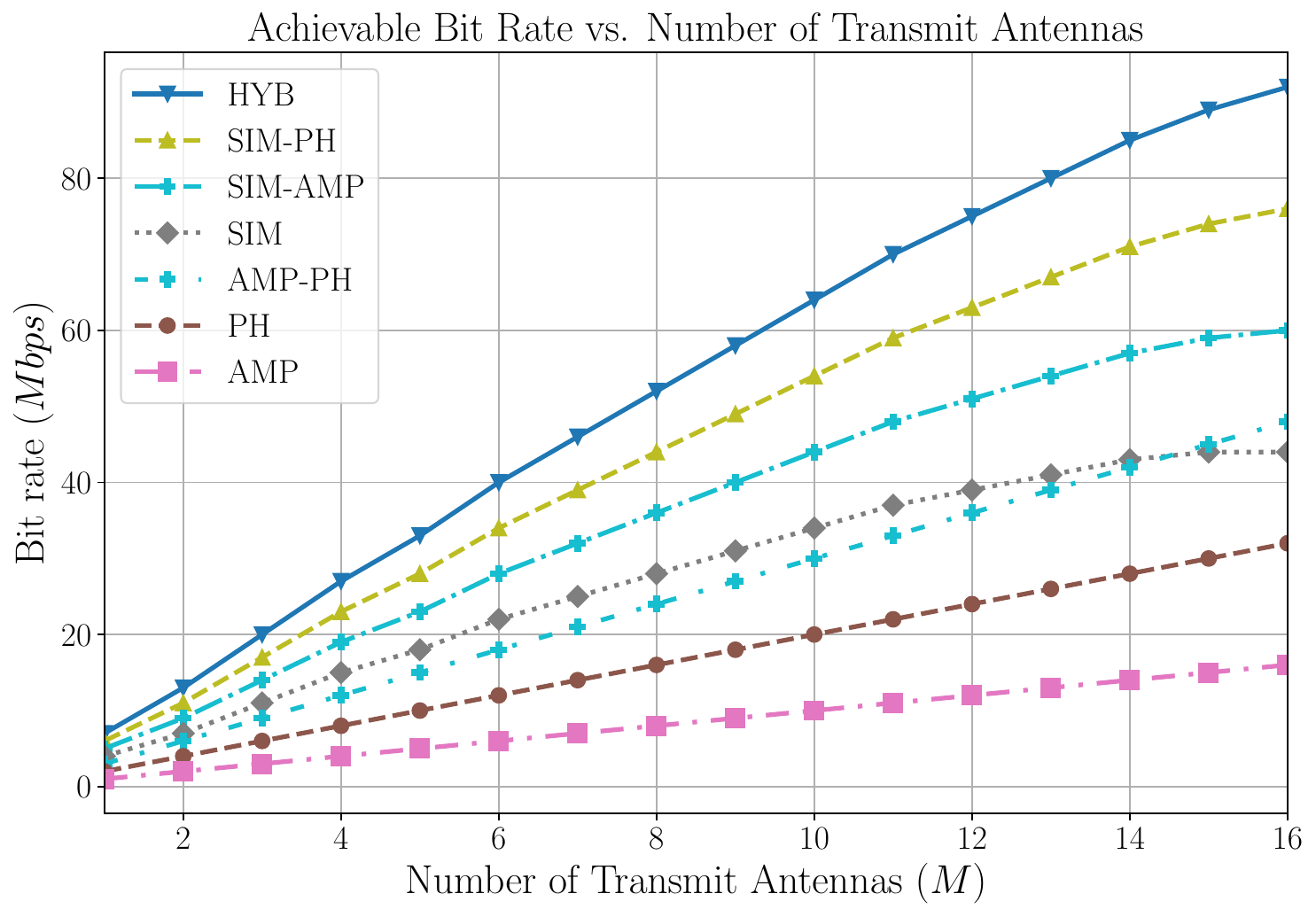}
\caption{Achievable bit rate as a function of the number of transmit antennas ($M$) for various embedding schemes.}
\label{fig:Bit_Rate}
\end{figure}

Figure~\ref{fig:Bit_Rate} compares the achievable rates of different schemes as a function of $M$ for $K=16$. 
Both \ac{AMP} and \ac{PH} embedding scale linearly with $M$, though \ac{AMP} achieves lower rates when $J_{ASK}<J_{PSK}$.  
In contrast, spatial index modulation (\ac{SIM}) yields higher rates than \ac{AMP} and \ac{PH}, with logarithmic growth. 
The rate increases as $M$ grows from $1$ to $K/2$, then decreases for $M>K/2$ due to the behavior of $\binom{K}{M}$. Thus, selecting $M=K/2$ provides a balance between rate and antenna utilization.  
And trivially, the hybrid (HYB) method achieves the highest rate, making it the most suitable for high-throughput \ac{ISAC} applications.

\subsection{Bit Error Rate and Secrecy Analysis}
\begin{figure}[t]
\centering
\includegraphics[width=\columnwidth]{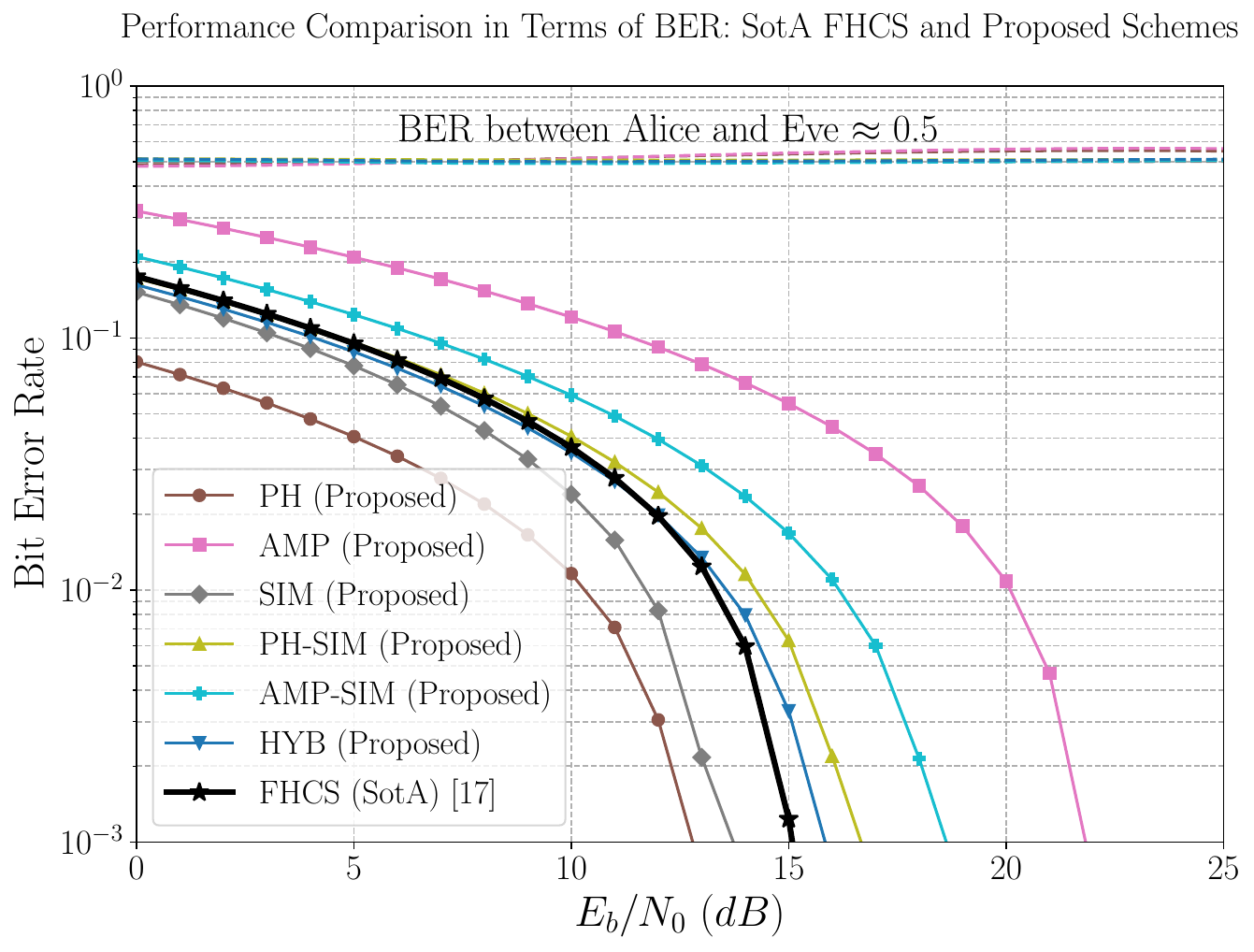}
\caption{\ac{BER} vs. $E_b/N_0$ for Alice-Bob (A\&B) and Alice-Eve (A\&E) links across SotA \ac{FHCS} and the proposed embedding schemes.}
\label{fig:Bit_Error_Rate}
\end{figure}
The \ac{BER} quantifies the proportion of bits in the key generated between Alice and Bob that are mismatched.
From Eve's perspective, a \ac{BER} close to $50\%$ reflects optimal security \cite{AldaghriTIFS_2020}. 
Fig.~\ref{fig:Bit_Error_Rate} presents the \ac{BER} performance for Alice-Bob and Alice-Eve links across different embedding schemes under varying \ac{SNR}. 
For all schemes, Eve's \ac{BER} remains near $0.5$ regardless of \ac{SNR}, since she lacks knowledge of the secret sequences $\mathbf{\Gamma}_T$ and $\mathbf{\Gamma}_f$ and must resort to random guessing. 
This limits her information gain and enhances privacy. 
In contrast, Alice and Bob achieve reliable communication with \ac{BER} approaching zero for $E_b/N_0$ values up to about 22~dB.  
This robustness is enabled by the use of matched filtering and \ac{OMP} at the receiver, both known for high noise resistance. 
The \ac{PH} scheme offers particularly strong performance, benefiting from identical $\mathbf{\Gamma}_T$ and $\mathbf{\Gamma}_f$ sequences at transmitter and receiver. 
Likewise, \ac{SIM} performs well since no information is embedded in chip amplitudes or phases, and the 1-sparse \ac{OMP} receiver effectively mitigates noise in the frequency domain. 
By contrast, \ac{AMP} is more noise-sensitive, as amplitude variations are more easily distorted than phase shifts.  
The \ac{BER} of \ac{PH}-\ac{SIM} and HYB schemes is similar and depends on parameters such as $J_{PSK}$, $K$, $M$, and $Q$. 
For instance, larger $J_{PSK}$ may increase the \ac{BER} of \ac{PH}-\ac{SIM}. 
Errors from \ac{SIM} also propagate into \ac{PH}-\ac{SIM} and \ac{AMP}-\ac{SIM}, explaining their relatively weaker performance. 
However, HYB outperforms \ac{AMP}-\ac{SIM} due to the stabilizing effect of the \ac{PH} component.  

The figure also highlights that our sparse MF receiver efficiently recovers the HYB signal, achieving a comparable rate and nearly identical $E_b/N_0$ performance to the \ac{SotA} \ac{FHCS} scheme. 
Unlike \ac{FHCS}, which requires an exhaustive search over all $\binom{K}{M}$ index combinations and $M!$ permutations \cite{BaxterTSP_2022}, our receiver significantly reduces computational complexity while preserving performance.  
Analytically, the secrecy rate in the wiretap channel can be defined as
\begin{equation}
    C_s = \max_{P(\mathbf{X}^{\mathrm{(Alice)}})} \left[ I(\mathbf{X}^{(\text{Alice})}; \mathbf{R}^{(\text{Bob})}) - I(\mathbf{X}^{(\text{Alice})}; \mathbf{R}^{(\text{Eve})}) \right]^+,
\end{equation}
where mutual information $I$ is inversely related to \ac{BER}. 
Thus, a low \ac{BER} between Alice and Bob increases $I(\mathbf{X}^{(\text{Alice})}; \mathbf{R}^{(\text{Bob})})$, while Eve's high \ac{BER} (ideally 0.5) reduces $I(\mathbf{X}^{(\text{Alice})}; \mathbf{R}^{(\text{Eve})})$, thereby improving secrecy and ensuring secure communication.

\subsection{Ambiguity Function}
\label{subsec:AF_Evaluation_Results}

Finally, the effectiveness of the radar sensing function is evaluated through the \ac{AF}, which is shown to remain even with the proposed embedding and security enhancements.

\begin{figure}[h]
\vspace{-3ex}
\centering
\includegraphics[width=\columnwidth]{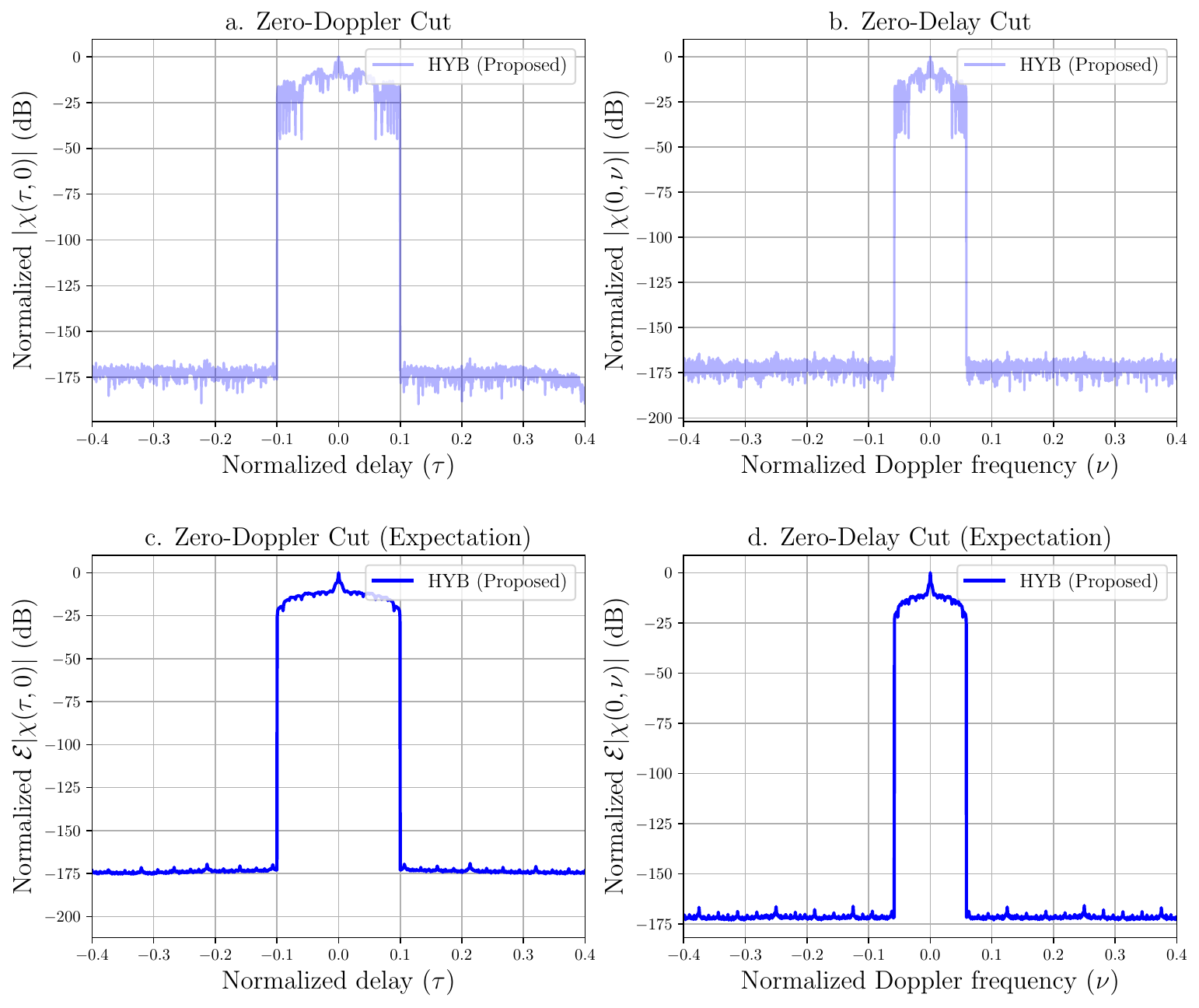}
\vspace{-1.5ex}
\caption{{Comparison of the Ambiguity Function (AF): (a) Zero-Doppler cut for various schemes, (b) Zero-delay cut for proposed schemes, (c) Zero-Doppler cut of the AF expectation for proposed schemes.}}
\label{fig:AF_plots}
\end{figure}

Fig. \ref{fig:AF_plots} compares the \ac{AF} of our proposed hybrid scheme.
The zero-Doppler cut (range profile) in Fig. \ref{fig:AF_plots}(a) shows a narrow mainlobe with low sidelobes, which is crucial for accurate target detection and effective clutter suppression.
Similarly, the zero-delay cut (Doppler profile) in Fig. \ref{fig:AF_plots}(b) demonstrates superior velocity resolution. 
In conventional fast-time embedding, varying FH codes typically raise sidelobes and degrade suppression performance. To address these challenges, we designed ISAC waveforms using \ac{RFPA} schemes, where randomized pulse repetition intervals and frequency variables jointly enhance range resolution, clutter robustness, and velocity estimation accuracy. The expectation of the \ac{AF}, shown in Fig.~\ref{fig:AF_plots} (c-d), then highlights a clean and sharp mainlobe, illustrating the benefits of this randomization.

\section{Conclusion}
\label{Conclusion}

We presented a novel framework for secure and high-performance \ac{ISAC}, with a design of an \ac{RFPA} waveform that employs shared secrets to randomize temporal and spectral properties, ensuring resilience against passive eavesdroppers. 
In addition, we introduced a hybrid modulation scheme over the \ac{RFPA} waveform, with a dedicated low-complexity sparse receiver to enable practical decoding without degrading \ac{BER} performance.  
Simulation results and \ac{AF} analysis verified that the proposed framework achieves strong security guarantees, enhances communication throughput, and improves radar sensing resolution compared to existing \ac{FH}-based \ac{ISAC} methods. 

\vspace{-1ex}

\section*{Acknowledgment}

This work was funded by the German Federal Ministry of Education and Research (No. 16KISK231), the German Research Foundation (Germany's Excellence Strategy-EXC2050/1-ProjectID 390696704-Cluster of Excellence CeTI of Dresden, University of Technology), and based on the budget passed by the Saxon State Parliament.

\vspace{-1ex}


\end{document}